\documentclass[aps,epsfigure,twocolumn,superscriptaddress,showkeys,nofootinbib]{revtex4-1}
\usepackage[colorlinks=true,linkcolor=blue,urlcolor=blue,citecolor=blue,pdfusetitle]{hyperref}
\usepackage[utf8]{inputenc}
\usepackage[english]{babel}
\usepackage{amsmath}
\usepackage[caption = false]{subfig}

\usepackage{graphicx,epstopdf}
\usepackage{blindtext}
\usepackage[table,xcdraw]{xcolor}
\usepackage{lipsum}
\usepackage{amsfonts}
\usepackage{bbm}
\usepackage{amssymb}
\usepackage{enumerate}
\usepackage{color}
\usepackage{latexsym}
\usepackage{times,txfonts}
\usepackage[normalem]{ulem}
\graphicspath{ {./figures/} }

\usepackage{amsmath}

\begin{document}

\title{Evolutionary-enhanced quantum supervised learning model}

\author{Anton Simen}
\email{anton.simen@kipu-quantum.com}
\affiliation{Kipu Quantum GmbH, Berlin, Germany}
\affiliation{SENAI CIMATEC, Salvador, BA, Brazil}

\author{Rodrigo Bloot}
\email{rgbloot@gmail.com}
\affiliation{Federal University of Latin-American Integration, Foz do Iguaçu, PR, Brazil}

\author{Otto M. Pires}
\email{otto.pires@fbter.org.br}
\affiliation{SENAI CIMATEC, Salvador, BA, Brazil}

\author{Erick G. Sperandio Nascimento}
\email{erick.sperandio@surrey.ac.uk}
\affiliation{SENAI CIMATEC, Salvador, BA, Brazil}
\affiliation{Surrey People-Centred AI Institute, University of Surrey, Guildford, Surrey, UK}

%%%%%% Affiliations %%%%%%

%%%%%% Date %%%%%%
% Date is optional
%\date{\today}

%%%%%% Abstract %%%%%%
\begin{abstract}

% Quantum supervised learning models based on variational circuits has been shown to be one of the most promising technologies that can be run on NISQ devices, thanks to its characteristics that demand few hardware resources, both in the creation of quantum feature maps and in the hardware-efficient ansatz containing trainable parameters. However, one of the bottlenecks faced in the training of quantum models is the barren plateau phenomenon, whose main consequence is the stagnation of learning throughout the iterations of the optimizers. This work introduces an evolutionary-enhanced ansatz-free supervised learning model, which takes advantage of quantum feature maps and, instead of a parametrized circuit, uses circuits with variable topology that evolves based on an elitist method. Also we introduce  a  \textit{superposition of multi-hot encodings} enabling the treatment of multi-classification problems. The introduced framework is able to avoid barren plateaus increasing the accuracy of the models. The results show a significant improvement in the training and precision of the models when compared to the  Variational Quantum Classifiers that are presented in the state-of-the-art of the technology. Furthermore, tests were conducted with a specific dataset class that conventional kernel machines struggle to efficiently separate, showing an alternative path towards quantum advantage on supervised learning.

Quantum supervised learning, utilizing variational circuits, stands out as a promising technology for NISQ devices due to its efficiency in hardware resource utilization during the creation of quantum feature maps and the implementation of hardware-efficient ansatz with trainable parameters. Despite these advantages, the training of quantum models encounters challenges, notably the barren plateau phenomenon, leading to stagnation in learning during optimization iterations. This study proposes an innovative approach: an evolutionary-enhanced ansatz-free supervised learning model. In contrast to parametrized circuits, our model employs circuits with variable topology that evolves through an elitist method, mitigating the barren plateau issue. Additionally, we introduce a novel concept, the "superposition of multi-hot encodings," facilitating the treatment of multi-classification problems. Our framework successfully avoids barren plateaus, resulting in enhanced model accuracy. Comparative analyses with Variational Quantum Classifiers from the technology's state-of-the-art reveal a substantial improvement in training efficiency and precision. Furthermore, we conduct tests on a challenging dataset class, traditionally problematic for conventional kernel machines, demonstrating a potential alternative path for achieving quantum advantage in supervised learning for NISQ era.

%Motivação (problema científico a ser abordado) 1 (uma) ou 2 (duas) frases
%Próposito - Objeivo do artigo  - 1 (uma) ou 2 (duas) frases
%Metodologia - 1 (uma) frase
%Resultado apresentado - 1 (uma) frase
%Conclusão - 1 (uma) frase
\end{abstract}
%%%%%% Main Text %%%%%%
\maketitle

\section{Introduction}
The theoretical concept of quantum computing is consolidated  and the fundamental theory is well established (see, e.g., \cite{Mm1}). However, the construction of a functional large scale universal quantum computer is still far from practical applications. Nowadays, some hardware proposals have been presented and, there are even commercial models available. Such devices can be classified as Noisy Intermediate-Scale Quantum (NISQ) (see, e.g., \cite{Preq}) because its intermediate scale operating with at maximum number of few hundred qubits. For such a kind of hardware, the variational quantum algorithms described in \cite{vqa} are suitable to be used on such devices in a certain range of tasks including spectral analysis of molecules, principal component analysis and graph partitioning (see, e.g., \cite{Anton}). The potential to good performance in machine learning arises as a consequence. In special, in supervised learning applications.

Supervised learning methods use data samples based on previous observations to train a model capable to make prediction about unseen samples. One of the most famous supervised learning methods is the Support Vector Machine (SVM) with formulation described in \cite{csvm}. The SVM method can be formulated as a quadratic programming problem representing an advantage in comparison to other kernel-based methods. By the other hand, the major drawback is related to data which are hard to separate even for higher dimension feature spaces.
 Among the difficult tasks to
accomplish good results using classic SVM we can cite the
non trivial process to make a prediction on three-dimensional
protein structures from a given data set where is specified the
protein sequence ( see, e.g., \cite{Jerbi} and\cite{Jump2021} ).

The Quantum Support Vector Machine (QSVM) method for real scenarios was initially proposed by \cite{Loid}, demonstrating a runtime speed-up of $\mathcal{O}(\log{FT})$ (with $F$ as the dimension of the feature space and $T$ as the number of training vectors) during both training and test procedures. However, for this method to work effectively, a coherent superposition framework is necessary, as emphasized by \cite{qsvm}. They presented a formulation capable of handling training and test data provided classically.

Considering the insights from reference \cite{qsvm}, it's noteworthy for contributing to a better understanding of QSVM. The authors suggest a pathway to achieve quantum advantage, particularly in cases where obtaining the kernel function classically is challenging. However, the practical application of these procedures to real-world problems remains unclear at this point.

A rigorous  speed-up for supervised learning was introduced in \cite{Yunchao} and considering just the case where the data is provided in a classical way consequently elucidating the questions raised by \cite{qsvm}. However, the main assumption used by the authors is the classical hardness of discrete logarithm problem used for the purposes of binary classification. 

One major challenge in implementing quantum supervised learning algorithms is the issue of the "barren plateau," which refers to a situation where the optimization landscape of the quantum circuit exhibits little gradient information, leading to slow or ineffective optimization. Strategies to mitigate the effects of the barren plateau, such as designing the quantum circuit with a more structured layout or using specialized optimization algorithms, are an active area of research. Since its inception, the concept of the barren plateau has gained widespread attention in the quantum computing community \cite{Jarrod}, with many researchers attempting to understand its underlying causes and develop strategies to mitigate its effects. In particular, the issue of the barren plateau has been extensively studied in the context of quantum kernel models \cite{cerezoP}, where it has been shown to be a major bottleneck for the performance of these algorithms.

 Besides of previously mentioned, other several authors have been working with quantum machine learning in the last years. Among them the prominent ones were compiled in the self contained comprehensive tutorials given by  \cite{Schud2022} and \cite{schudbook}. As well as  the choice of method used for better recognition rate is fundamental, encoding data sets is another important task in supervised quantum learning which can be found with good details in \cite{Schud2021}, \cite{Malacuso}, \cite{Casper} and \cite{Araujo}. The latter proposed an interesting approach by using  divide‑and‑conquer algorithm for quantum state preparation  successfully tested on real devices.

In this work a quantum supervised learning model is proposed following similar lines given in \cite{Circlern0}. However, here the circuit training is performed through an evolutionary procedure introduced in \cite{qce} and adapted for our framework. Through empirical computational experiments, we show that this evolutionary algorithm, in conjunction with the already known quantum feature maps, presents performance gains on multiclassification as well as on binary classification tasks when compared to its variational counterpart. More specifically, experiments were carried out comparing the proposed approach named Evolutionary Quantum Classifier (EQC) with the Variational Quantum Classifier (VQC). Experiments were also conducted with ad-hoc datasets, where conventional kernel machines cannot effectively separate unless quantum-inspired kernel functions are created – a task that proves inefficient with high-dimensional data. These findings underscore the need for a more comprehensive exploration of real-world data sources, such as those found in \cite{qm91, qm92} dataset, which may exhibit similar structures to those used in our tests. Such data could be invaluable in addressing challenges like molecular property prediction, drug discovery, and more (see, e.g.,\cite{qm9_1}).

% In this paper the speed-up or quantum supremacy (or advantage) is not the main target. Based on this, the considered data is the iris database (multiclass) and the other two are artificial (binary) data specially prepared so that the classic classifier has worse performance. The main objective is to compare the performance between quantum classifiers in two very different scenarios. In the first scenario, with regard to iris, the classical technique demonstrates robustness in relation to its quantum counterparts. However, the effects of the phenomenon known as barren plateaus can be seen in the variational classifier in opposition to the evolutionary one, which appears to be free of this effect. In the second scenario, where the quantum approaches are favorites for best performance, we can see again the better performance of the evolutionary framework.

\section{Quantum Feature Map}
We start presenting a brief summary about the quantum counterpart to the classical kernel methods.
Let the attributes be ${\bf x} \in X$, where $X$ ($X\subset \mathbb{R}^n$) represents a non-linearly separable dataset of dimension $n$ with $k$ classes. In order to map the data to a space where it becomes linearly separable by a hyperplane, kernel methods are used. From the mathematical definition of a kernel function, we have the kernel function $K({\bf x},{\bf z}) = \langle \phi({\bf x}), \phi({\bf z}) \rangle $, ${\bf x}$ and ${\bf z}$ being \textit{n}-dimensional vectors. The $\phi({\bf x})$ function maps $\phi:X \subset \mathbb{R}^n \rightarrow \mathbb{R}^m$, where $m$ is usually much larger than $n$. 
In the quantum counterpart we can build the map function in an infinite dimensional space. As a consequence, the
 quantum feature map is a function that strictly plays the same role such that the mapping is done using an \textit{n}-qubit operator so the result is a vector that lives in the higher-order Hilbert space, $\mathcal{H}$. So $\Phi: X \rightarrow \mathcal{H}$. The quantum  general form feature map has the expression
\begin{equation}
    \mathcal{U}_{\Phi({\bf x})}|0\rangle^{\otimes n} = |\Phi({\bf x})\rangle,
    \label{map1}
\end{equation}
which has been shown to play an important role toward the quantum advantage on supervised learning tasks on quantum computers for specifics data sets \cite{speedupsupervised}. The unitary operator $\mathcal{U}_{\Phi(x)}$ is built using the unitary operator ${V}_{\Phi(x)}$ combined with Hadamard  gates as follows
\begin{equation}
    \mathcal{U}_{\Phi({\bf x})}={V}_{\Phi({\bf x})}H^{\otimes n}{V}_{\Phi({\bf x})}H^{\otimes n}.
    \label{map2}
\end{equation}

\begin{figure}
    \centering
   \includegraphics[width=0.37\textwidth]{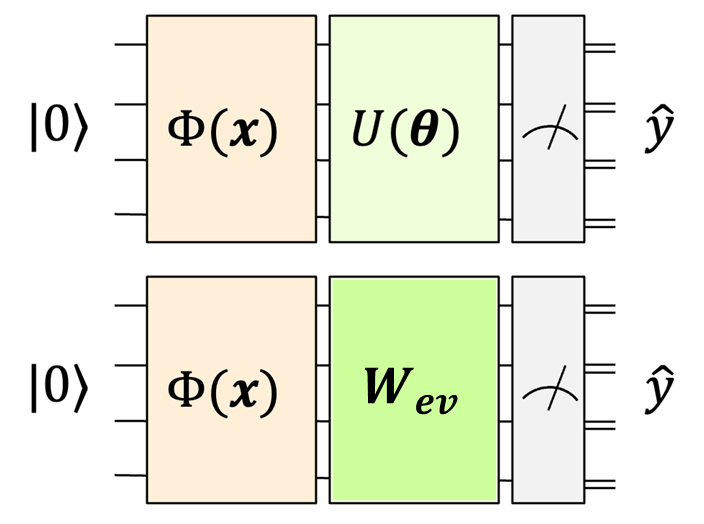}
    \caption{Kernel-based quantum circuits for multi-class classification: the quantum feature map block is the operation which maps classical data into a higher-order Hilbert space. Given above, the conventional explicit quantum model. The evolutionary circuit block (illustrated below) evolves through single and two-qubit unitary operation based on an elitist method. The number of measured qubits depends on the type of label encoding as well as the number of labels.}
   \label{fig:1}
\end{figure}

The unitary operator $\mathrm{V}_{\Phi({\bf x})}$ is defined in order to ensure that the process cannot be easily reproduced on classical computers for large instances.
Furthermore, in a different way used in \cite{qsvm}, a quantum feature map can be also defined as expression $\mathcal{U}_{\Phi({\bf x})} = \bigotimes_{i=0}^{n}RX(x_i)$, since ${\bf x}$ should be normalized using min-max approach such that $x_{i} \in \left[0, 2\pi\right)$. Such quantum feature map does not produce any effect in circuit depth, remaining $\mathcal{O}(1)$. An arbitrary representation of a feature map is illustrated in Fig. \ref{fig:1}, combined with a trainable quantum circuit.
%--------------------------------------------------------------------
\section*{Quantum Circuit Evolution  Algorithm }

The Quantum Circuit Evolution Algorithm (QCE)  proposed by \cite{qce}, in contra-position to the ansantz-dependent variational methods, is ansatz-free and perform updates in all circuit configuration. Therefore, using an evolutionary scheme at each generation the circuit is adapted to optimize the cost function which will now be circuit-dependent. In other words, we have now a cost function $F$ given by 
\vspace{6pt}
\begin{center}
\begin{equation}
 F(\mathbf{C})=\langle\psi_0|\mathbf{C}^{\dagger}\mathcal{H}\mathbf{C}|\psi_0\rangle,
\label{eq10}
\end{equation}
\end{center}
\vspace{6pt}
where $\mathbf{C}$ is a circuit, $\mathcal{H}$ (in the context of the present paper) is an observable and $|\psi_0\rangle$ is an arbitrary initial state. The circuit is represented by an unitary operator indexed by $d$ given by the circuit $\mathbf{C}=\mathbf{U}_{d}\mathbf{U}_{d-1}\cdot\cdot\cdot\mathbf{U}_{1},$ which we want to determine in order to optimize the cost function $F$. The technique works by using rotation operators $U_{k}=U(L_{k},\theta_{k})$ where $L_k$ is obtained from the set of gates $\{R_x,R_y,R_z,R_{xx}, R_{yy}, R_{zz}\}$ which are generated from the canonical basis $\{\sigma_x,\sigma_y,\sigma_z\}^{\otimes 2} $. Finally (for a fixed $\theta_k$) the explicit form of $U_{k}=U(L_{k},\theta_{k})$ is defined as
\begin{equation}
    U(L_{k},\theta_{k})=exp\left[-i\frac{\theta_k}{2}L_{k}\right].
    \label{circ}
\end{equation}
 
  In the case of the variable circuit topology, from an initial random population a mutation strategy is performed to obtain the best circuit $\mathbf{C}_{best}$ that optimizes the equation (\ref{eq10}). The actions which can be performed into the circuit are described in Fig. \ref{fig:3}.
 
 % \begin{itemize}
 %     \item  Add random gate at a random position;
 %     \item Delete gate at a random position;
 %     \item Replace gate at a random position with a random new gate;
 %     \item  Change parameter of a gate at a random position.
 % \end{itemize}
 % 1) Add random gate at a random position; 2) Delete gate at a random position; 3) Replace gate at a random position with a random new gate; 4) Change parameter of a gate at a random position. 
 The procedure used in the present paper follows, with some modifications, the approach used in \cite{qce} and, consequently, at the end of circuit-generation evolution we have
 \vspace{6pt}
\begin{center}
\begin{equation}
 \mathbf{C}_{best}=arg \min_{\mathbf{C}\in { \mathbb{G}}} F(\mathbf{C}),
\label{eqbest}
\end{equation}
\end{center}
\vspace{6pt}
where the set $\mathbb{G}$ is the search space which can be continuous or discrete.
 %-------------------------------------------------------------------
 
\textbf{Remark:} Methods based on fixed circuit topology are restrict to fixed circuit configuration and optimize the parameter ${\theta_k}$, which is a disadvantage as the circuit depth does not change during the process resulting in barren plateaus effect for several configurations of the cost functions (see, e.e., \cite{cerezoP}).
\section{Quantum Evolutionary Classification}

\begin{figure}
    \centering
   \includegraphics[width=0.3\textwidth]{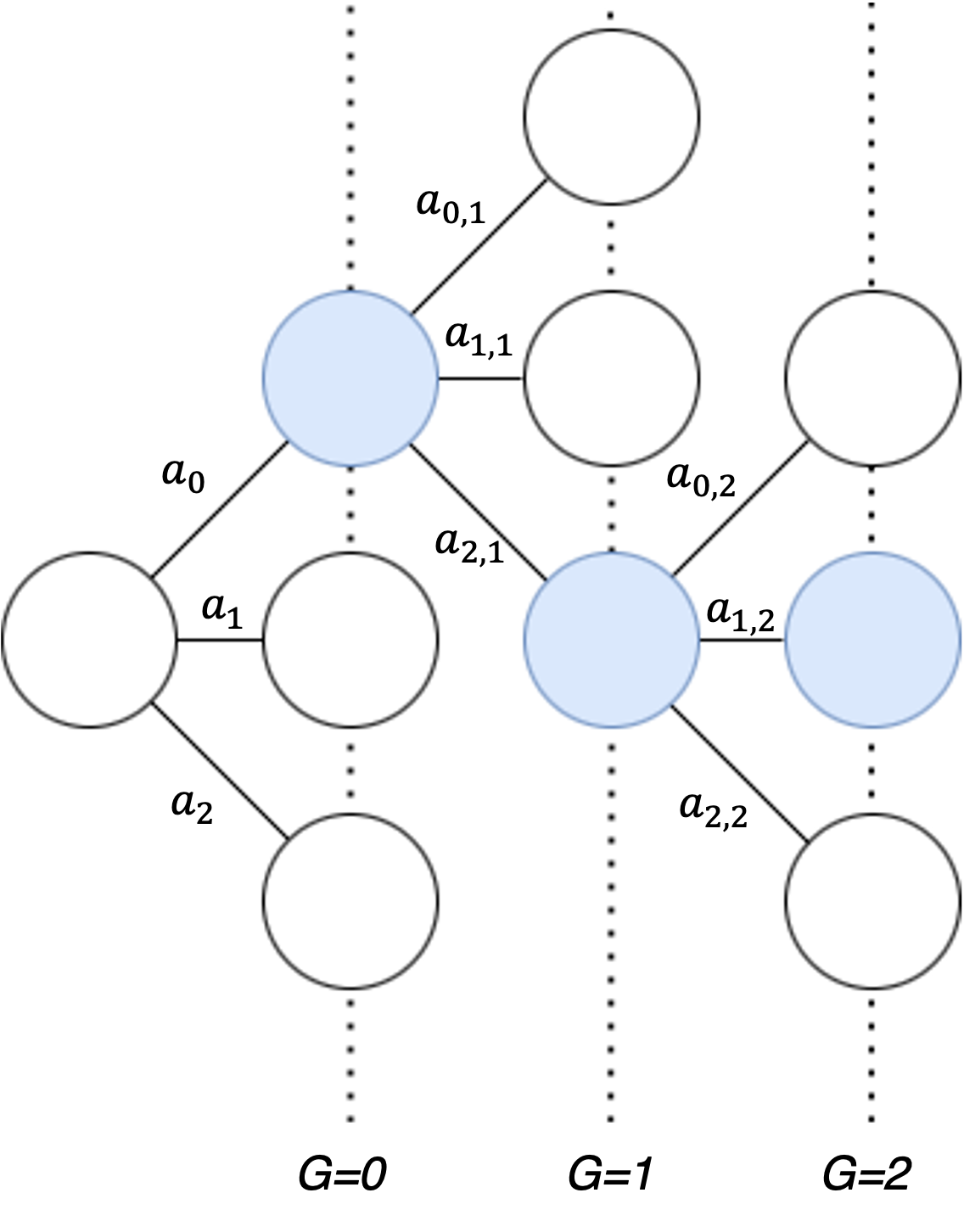}
    \caption{The tree shows how the elitist mechanism behaves over the generations. The quantum circuits (represented by nodes) are branched and actions, $a_{i,j}$, (\ref{fig:3}) are perfomed on this circuits. The elitist scheme select the circuit which produces the best result (blue nodes) and carry it out over the next generations $G$. }
   \label{fig:2}
\end{figure}

\begin{figure}
    \centering
   \includegraphics[width=0.4\textwidth]{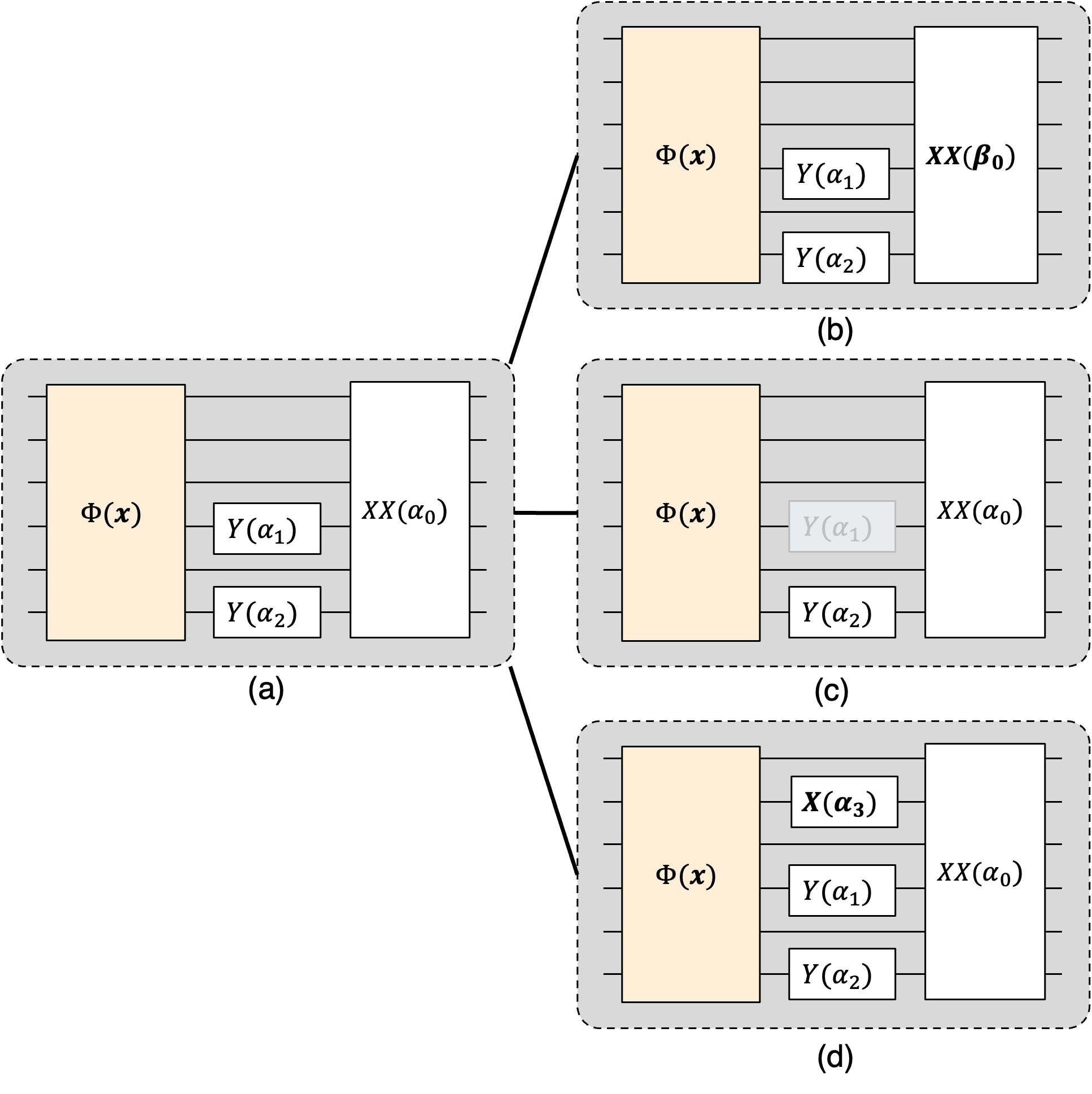}
    \caption{A schematic description of potential actions to the current circuit. (a) illustrates the current circuit diagram within a given generation. The subsequent figures below illustrate plausible actions that can be applied to the existing circuit. Action (b) modifies the rotation angle of the an existing gate. It is also viable to perform actions such as (c) deleting a gate which exhibits a negative influence on the objective function and (d) inserting a new gate, selected from a pool of single and multi-qubit gates. Note that, in this illustration, the block $XX(\alpha)$, represents the two-qubit unitary $e^{-i\alpha X_iX_j}$, acting on the $i$-th and $j$-th qubits (for instance, the first and last ones, respectively). Note that, for this illustration, $X(\alpha)$ is not related to the dataset $X$.}
   \label{fig:3}
\end{figure}

The evolutionary classification protocol follows a very similar structure to the one proposed in \cite{qsvm}. The key difference is that instead of implementing a layered ansatz operator, $U(\Vec{\theta})$, whose variational parameters, $\Vec{\theta}$, are updated iteratively, one apply the evolutionary operator, $\mathcal{W}_{ev}$, whose evolution is based on an elitist method introduced in \cite{qce}, which was originally proposed to solve combinatorial optimization problems with a demonstration of improving performance and as a good option to deal with inherent vanishing gradients of variational algorithms. The evolutionary operator $\mathcal{W}_{ev}$, starts with a quantum circuit depth $d=1$ then evolves at each generation until minimize a loss function, $\mathcal{L}(\hat{y}, y)$, which measures error between an estimator, $\hat{y}$, and its respective label, $y$, from training set. Both approaches are illustrated in a general way in Fig. \ref{fig:1}, on the other hand the evolutionary approach is described in details in Figs. \ref{fig:2} and \ref{fig:3}. 

The loss function can be calculated and chosen according to the number of labels in the data set, and can be for binary or multi-label classification models. In both cases, the loss function is calculated from the probability distribution measured in the state 
\begin{equation}
|\Psi(x)\rangle = \mathcal{W}_{ev}\mathcal{U}_{\Phi(x)}|0\rangle^{\otimes n}.
\label{psi_x}
\end{equation}

\subsection{Binary classification}
For a dataset with two labels, $y \in \{-1, +1\}$, the estimator is obtained by measuring qubits in computational $z$-basis using a parity function, $p(x)$, given by the expected value of an observable, $P = \bigotimes_{i=0}^{n}b_i$, where $b_i \in \{\sigma_z, I\}$.  For example, for a two-dimensional data with two classes, a parity function could be given by $p(x)=\langle \sigma_z \otimes \sigma_z \rangle$, if all the qubits are measured. For binary classification it is also possible to measure a single qubit, being $p(x)=\langle \sigma_z\rangle$ the parity function. Thus, the loss function can be given by

\begin{equation}
    \mathcal{L} = \frac{1}{|X|}\sum_{((\Vec{x}),y)\in X} l(p(\Vec{x}), y)
\end{equation}
where $p(x)$ can be rewritten explicitly as $p(x) = \langle \Psi(x)|P|\Psi(x)\rangle$. There are multiple options for evaluate the loss function, $l(p(x), y)$, such as \textit{Mean Squared Error} (MSE) and \textit{Log-loss}.

\subsection{Multiclass classification}

For data sets whose number of labels are greater than two, other strategies must be adopted although it is similar to binary classification. The label encodings adopted in this work are based on superposition of eigenstates of the computational basis, which we will call \textit{superposition of multi-hot encodings}. Given a probability distribution, $\Omega$, obtained from measurements of the quantum circuit in the $z$-basis, the estimators, $\hat{y}_i$, in this case, are calculated as sums of disjoint subsets in $\Omega$ , being chosen empirically, distributing such probabilities uniformly among labels from $X$. As before mentioned,  $X$ is a dataset of dimension $n$ with $k$ classes. Using a quantum feature map that requires $n$ qubits, $\Omega$ would have a maximum dimension of $2^n$ and it would be possible to choose different encodings for the labels of $X$. For a dataset with $n$ dimensions and $k$ classes, where $Y = \{y_0, y_1, y_2, ..., y_k\}$, the discrete probability distribution after measurements on the state $|\Psi(x)\rangle = \sum_{i=0}^{2^n}\alpha_i|i\rangle$ is given by $\Omega=\{\ \|\alpha_0\|^2,\|\alpha_1\|^2,\|\alpha_2\|^2, ..., \|\alpha_{2^n}\|^2 \}$. A choice for the encoding would be a summation of eigenstate probabilities, given by $\Omega$, evenly divided for each of the $k$ classes. Then, the estimators can be defined as
\begin{equation*}
     \hat{y}_j = \sum_{i \in \Omega_j} \|\alpha_i\|^2,
\end{equation*}
where $\Omega_j$ is a disjoint subset of $\Omega$. Note that the probabilities used to calculate each estimator are projections on the $z$-basis given by $\|\alpha_i\|^2 = |\langle \Psi(x) | i\rangle |$. It is noteworthy that any other choice for encoding would be valid according to the chosen observable $\mathcal{H}$. However, the results that will be presented in this paper show that the approach chosen here has acceptable performance in the studied databases. Finally, the size of $\Omega$ must be sufficiently larger than the number of $k$ classes, depending on the encoding type.

% \begin{figure}
%     \centering
%    \includegraphics[width=0.35\textwidth]{figures/correlation_map_iris.png}
%     \caption{Heat map representing the feature correlation on Iris data set: the feature acronyms SL, SW, PL and PW stands for sepal length, sepal width, petal length and petal width, respectively.}
%    \label{fig:1}
% \end{figure}
\section{On Barren plateaus effect}
 As illustrated in \cite{Jarrod}, Barren plateaus is a intrinsic problem related when it is considered  random parameterized quantum circuits (RPQCs) with a fixed topology. An option to overcome this effect was proposed in \cite{cerezoP}. In this paper the authors propose the formulation of cost functions using a local observable instead of global ones. In the case of circuit evolution, the parameter is fixed in relation to the circuit topology which, in turn, varies making possible to show (under the conditions adopted) that circuit evolution method is almost free from the barren plateaus effect. The cost function (considering for simplicity a circuit $\mathbf{C}$ with real values) can be written in the following form 
\begin{equation}
    F(\mathbf{C})=tr\left[\mathbf{C}|\psi_0\rangle\langle\psi_0|\mathbf{C}^{T} \mathcal{H}\right],
\end{equation}
where the derivative of cost function, with respect the circuit $\mathbf{C}$, has the expression
\begin{equation}
    \frac{\partial F(\mathbf{C})}{\partial\mathbf{C}}=\nabla_{\mathbf{C}}tr\left[\mathbf{C}|\psi_0\rangle\langle\psi_0|\mathbf{C}^{T} \mathcal{H}\right].
    \label{gradcirc}
\end{equation}

As mentioned before in the main text, the defined operator $\mathcal{H}= P=\bigotimes_{i=0}^{n}b_i$ is an observable here considered as a real diagonal operator with $\mathcal{H}^{T}=\mathcal{H}$. Also, there is no variation of $\vec{\theta}$ which is connected to the circuit choice. The derivative of cost function in relation to the circuit is given by
\begin{equation}
    \frac{\partial F(\mathbf{C})}{\partial\mathbf{C}}=\mathcal{H}\mathbf{C}|\psi_0\rangle\langle\psi_0|^{T}+\mathcal{H}\mathbf{C}|\psi_0\rangle\langle\psi_0|.
    \label{gradcirc2}
\end{equation}
In the context of the framework proposed in this paper, we consider $|\psi_0\rangle\langle\psi_0|=|0\rangle\langle 0|$ and, for the  worst case where the circuit $\mathbf{C}$ has nonzero elements only in the diagonal, one can show the expression:
\begin{equation}
\frac{\partial F(\mathbf{C})}{\partial\mathbf{C}} = 
\begin{cases}
    2\mathcal{H}_{lh}C_{lh}, & \text{if } l=h=1 \\
    0, & \text{otherwise}.
\end{cases}
\end{equation}
The only two possibilities to make the derivative of the cost function zero are $\mathcal{H}_{11}=0$ or/and ${C}_{11}=0$. The observable is given in terms of the tensorial product of Pauli gates $\{\sigma_z,I\}^{\otimes n}$ which implies $\mathcal{H}_{11}=\pm1$ and, as a consequence, the derivative matrix will be null only if ${C}_{11}=0$. However, each fixed parameter $\theta_j$ related to the to the built circuit is constant at the selected population and generation. It varies randomly in the range $[0,2\pi)$ which do not implies in exponential possibilities for the value of ${C}_{11}$ to be null. For the case where $\mathbf{C}$ is a real arbitrary circuit,  we have
\begin{equation}
    \frac{\partial F(\mathbf{C})}{\partial\mathbf{C}}=2\mathcal{H}\mathbf{C}|0\rangle\langle0|.
    \label{gradcirc5}
\end{equation}
In this case, the operator will be null only if the first column in the circuit  has all its values defined as zero. Therefore, given the variation in topology through mutations, there is no possibility of exponentially null results for the derivatives of the cost function for such case as well. In the case where $\mathbf{C}$ is complex the analysis is similar (see, e.g., \cite{dermat}). 
\section{Results}

The computational experiments were carried out using two different scenarios.
In the first scenario, the iris dataset (which is well known from the literature and it was first presented in \cite{Fisher}) was tested comparing the performance of the Support Vector Classifier (SVC) compared to the quantum classifiers. For this case, the Quantum Evolutionary Classifier (QEC) was compared with the Variational Quantum Classifier (VQC) in terms of test accuracy, number of epochs and barren plateau effect on convergence. 

In the second scenario, the test accuracy was compared between QEC, VQC and the SVC, running on the ad-hoc data set, generated from the exponentiation of Pauli operators. In the following, a detailed description of the two scenarios given  separately. For both experiment scenarios, the samples of each class have been partitioned in: $65\%$ for training and $35\%$ for test. Each subset has been randomly sampled, with no overlapping between the sets.

\subsection{First scenario: EQC and VQC compared with SVC}
% Lets start describing the results obtained in the first scenario which is given by the iris dataset. Such data has $150$ samples, being $50$ for each of the three species named, Iris setosa, Iris virginica and Iris versicolor. For this case was considered four attributes named features which are given by the length and width of sepals and petals. These measures are standard and are useful to create a linear discriminant model to classify the species. This database is suitable for presenting our point of view on the use of quantum algorithms in pattern recognition. Here, we perform a multiclass classification comparing the results obtained by the classical versus the quantum classifiers. 
Starting with an examination of results in the initial scenario, we delve into the iris dataset, encompassing 150 samples evenly distributed among three distinct species: Iris setosa, Iris virginica, and Iris versicolor, with 50 samples for each. Our analysis incorporates four essential attributes, specifically the dimensions of sepals and petals—comprising length and width. These standardized measurements form the basis for constructing a linear discriminant model tailored for species classification. The iris dataset serves as a pertinent illustration to articulate our perspective on the application of quantum algorithms in pattern recognition. Within this framework, we undertake a multiclass classification task, contrasting outcomes obtained through classical classifiers with those yielded by their quantum counterparts.

% For the classic classifier, the tests were made based in three kinds of kernels\footnote{https://scikit-learn.org/stable/}. The results, regarding to accuracy, for the three approaches are given by  Fig \ref{fig:5}. It is clear that, for this database, the linear-SVM and poly-SVM outperform the Gaussian Kernel. The polynomial kernel completely separated the data achieving perfect recognition.

The classical classifier was tested using three distinct kernel types\footnote{https://scikit-learn.org/stable/}. The accuracy results for these approaches are illustrated in Fig \ref{fig:5}. Notably, in the dataset context, the linear-SVM and poly-SVM exhibit superior performance compared to the Gaussian Kernel. The polynomial kernel stands out for achieving complete data separation, contributing to highly accurate recognition.

Exploring the quantum counterpart, let's dive into the details of how we encoded multi-class information for the examples under consideration. This process followed the steps outlined in the multi-class classification subsection. For this scenario, we utilized a probability distribution denoted by $\Omega=\{\ \|\alpha_0\|^2,\|\alpha_1\|^2,\|\alpha_2\|^2, ..., \|\alpha_{15}\|^2 \}$. As mentioned earlier, we defined distinct sets $\Omega_j$, where each set corresponds to a labeled class. The partitioning aimed at identifying the three classes is outlined as follows: $\Omega_1=\{\ \|\alpha_1\|^2,...,\|\alpha_5\|^2 \}$, $\Omega_2=\{\ \|\alpha_6\|^2,...,\|\alpha_{10}\|^2 \}$, and $\Omega_3=\{\ \|\alpha_{11}\|^2,...,\|\alpha_{15}\|^2 \}$ for Iris setosa, Iris virginica, and Iris versicolor, respectively. It's worth noting that we excluded $\|\alpha_0\|^2$ from this set partition selection. This specific partitioning configuration was chosen based on the observable algebraic structure.

For both the Evolutionary Quantum Classifier (EQC) and Variational Quantum Classifier (VQC), measurements were conducted using the Qiskit SDK with a quantum assembly language (QASM) simulator.

In the EQC implementation, specific hyper-parameters were selected: a $50\%$ probability for inserting a quantum gate circuit into the current circuit, a $30\%$ probability for modifying rotation angles of quantum gates in the current circuit, a $10\%$ probability for swap the target and control qubit for a gate in the current circuit, and a $10\%$ probability for deleting quantum gates. The individual update strategy employed was elitist, where $\mu = 4$ initial individuals (quantum circuits) were randomly generated based on the established probabilities for the hyper-parameters. The individual that optimizes the cost function is inherited by subsequent generations. In this study, a maximum of 500 generations was considered to observe the barren plateau phenomenon.

To maintain polynomial complexity for the proposed heuristic, the number of measures for the circuit was of polynomial order $\mathcal{O}(poly(N))$, where $N$ represents the number of qubits. This choice ensures scalability and efficiency in handling quantum computations.

In Fig \ref{fig:5}, it's evident that the classical counterpart outperforms both quantum approaches. Given the data residing in $\mathbb{R}^4$, where spatial class arrangement information is absent, the results from the classical linear kernel highlight nearly optimal separation in this higher dimension. Notably, employing a polynomial kernel achieves complete class separation with $100\%$ accuracy. In this study, default Gaussian parameters were used, but optimizing $C_{exp}$ and $\gamma$ parameters could enhance Gaussian performance.

\begin{figure}
    \centering
   \includegraphics[width=0.4\textwidth]{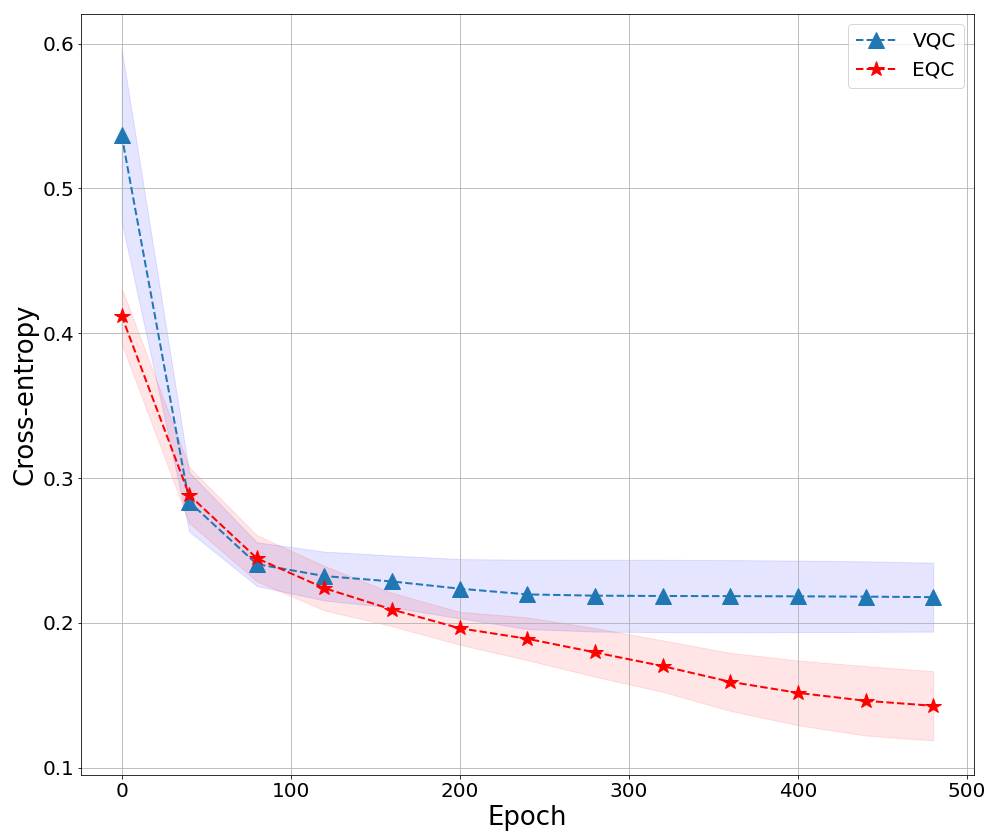}
    \caption{ Figure ilustrating the loss performance of the used quantum classifiers. In the case of EQC, for each new generation the new circuit changes its depth and parameters managing to escape from the plateau. Shadow plots represent the standart deviation over $10$ experiments.}
   \label{fig:4}
\end{figure}

Turning to quantum classifiers, as depicted in Fig. \ref{fig:5}, EQC demonstrates superior performance compared to VQC. As shown in \cite{Jerbi}, VQC is inherently linear, and EQC inherits this feature by construction. However, the barren plateau phenomenon significantly impacts VQC, contributing to its comparatively lower performance. This plateau effect on VQC outcomes is evident in Fig. \ref{fig:4}, underscoring the advantages of the EQC approach.

\subsection{Second scenario: EQC compared with VQC}
 The second scenario considers a synthetic generated dataset named 2-dim adhoc and 3- dim adhoc considering  two and three dimensions respectively. The artificial nature of these data is fundamental to make clear the
conditions where is preferable to use the quantum approach instead the classical classifier. The 2-dim adhoc is the very same introduced by \cite{qsvm} and it was considered under the same conditions. 

\begin{figure} 
    \centering
   \includegraphics[width=0.42\textwidth]{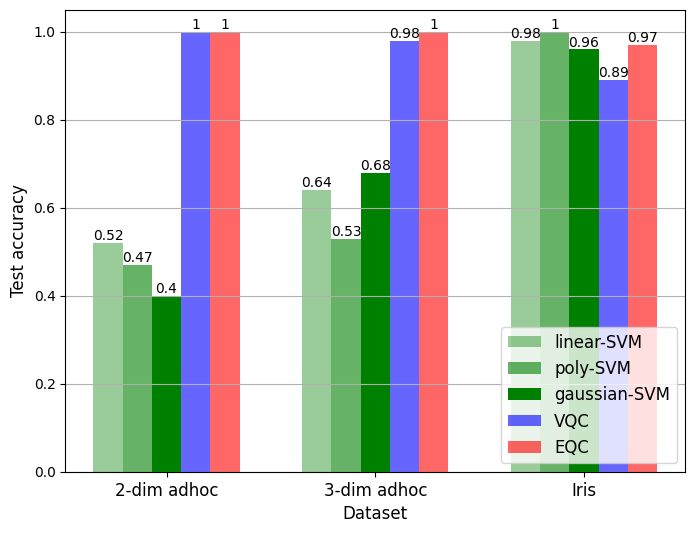}
    \caption{
Comparison of test accuracy across various models. The findings distinctly showcase the efficiency of the quantum approach in scenarios where the classical method lacks the potential for superior performance. Additionally, the results highlight the classical approach outperforming both quantum classifiers in the iris database.}
   \label{fig:5}
\end{figure}

The detailed description of how to obtain this data can be seen in that reference. Furthermore, there is no classical feature map capable to efficiently perform the complete separation on this data. The performance of classical kernels are represented in Fig. \ref{fig:5}, it is evident the downgrade on the obtained results in comparison to scenario one. However, it is clear the better performance of the quantum classifiers with both reaching a complete data separation. Also, no effect of barren plateau was observed on the VQC approach achieving the same performance of EQC for this data set. The same approach was used to create the 3-dim adhoc which can be seen using the qiskit tools. For this case, again the classical classifier is outperformed by the two quantum classifiers for all the tested classical kernels with the results illustrated in Fig. \ref{fig:5}. Regarding the comparison between quantum classifiers, we see that EQC obtained better classification results with $100\%$ of test accuracy. However, under model enhancement, it is also evidently expected the same accuracy for VQCs.

%\section{Discussion}
\qquad
\section{Comments and conclusions}
In this paper a new framework to pattern recognition using quantum algorithms was introduced. The framework uses an evolutionary circuits scheme to perform a supervised learning task and a named  \textit{superposition of multi-hot encodings} to deal with multi-classification. Several experiments were carried out in order to better understand the performance of the new approach. In order to make a fair comparison with classic methods we used the iris database in our experiments and an artificially generated data set where is expected that classical SVMs fail. Our major intention was to compare EQC and VQC for both data. For the iris data set, the classical counterpart outperforms both EQC and VQC, for the other side the performance of EQC for this data is much better than VQC.

 Circuit learning was introduced by \cite{Circlern0} with a framework based in parameter updating. As a consequence, the cost function is calculated  tuning the circuit parameters $\vec{\theta}$ in an iterative way. In a different approach, our framework uses an elitist strategy where the topology of the circuit is completely updated for each generation in this way changing also the circuit depth.
 
Machine learning algorithms with parameterized quantum circuits are proving useful in various pattern recognition tasks where classical methods face limitations due to the nature of the model. The EQC method, with its adaptable topology, contributes to a highly efficient accuracy rate. It's important to note, though, that while it excels, efficiency isn't guaranteed compared to classical counterparts.

Our experiments show cases where classical classifiers outperform quantum ones. In such instances, there's no reason to replace conventional methods with quantum ones. However, looking at complex examples highlights the need for quantum algorithms in real-world cases where classical approaches fall short.

In large-scale, realistic patterns, VQC's design can make it less scalable. Despite attempts to improve this using local observables, the gradient still decreases polynomially with the qubit count. In contrast, EQC, with its variable topology, proves effective in scaling scenarios where classical methods struggle. The topology variation significantly impacts cross entropy, improving learning performance with each generation update.

Our conclusion leans towards EQC's superiority due to its near-immunity to plateaus. This makes it a promising technique for scaling to a larger number of qubits in scenarios demanding quantum multi-classification. More research on hardware implementations is necessary, and our approach aligns well with the demands of the NISQ era. Also, encouraging additional research on real-world, intricate datasets that can be effectively separated using this approach is highly recommended.

\bibliography{sample}

\begin{thebibliography}{26}%
\makeatletter
\providecommand \@ifxundefined [1]{%
 \@ifx{#1\undefined}
}%
\providecommand \@ifnum [1]{%
 \ifnum #1\expandafter \@firstoftwo
 \else \expandafter \@secondoftwo
 \fi
}%
\providecommand \@ifx [1]{%
 \ifx #1\expandafter \@firstoftwo
 \else \expandafter \@secondoftwo
 \fi
}%
\providecommand \natexlab [1]{#1}%
\providecommand \enquote  [1]{``#1''}%
\providecommand \bibnamefont  [1]{#1}%
\providecommand \bibfnamefont [1]{#1}%
\providecommand \citenamefont [1]{#1}%
\providecommand \href@noop [0]{\@secondoftwo}%
\providecommand \href [0]{\begingroup \@sanitize@url \@href}%
\providecommand \@href[1]{\@@startlink{#1}\@@href}%
\providecommand \@@href[1]{\endgroup#1\@@endlink}%
\providecommand \@sanitize@url [0]{\catcode `\\12\catcode `\$12\catcode `\&12\catcode `\#12\catcode `\^12\catcode `\_12\catcode `\%12\relax}%
\providecommand \@@startlink[1]{}%
\providecommand \@@endlink[0]{}%
\providecommand \url  [0]{\begingroup\@sanitize@url \@url }%
\providecommand \@url [1]{\endgroup\@href {#1}{\urlprefix }}%
\providecommand \urlprefix  [0]{URL }%
\providecommand \Eprint [0]{\href }%
\providecommand \doibase [0]{http://dx.doi.org/}%
\providecommand \selectlanguage [0]{\@gobble}%
\providecommand \bibinfo  [0]{\@secondoftwo}%
\providecommand \bibfield  [0]{\@secondoftwo}%
\providecommand \translation [1]{[#1]}%
\providecommand \BibitemOpen [0]{}%
\providecommand \bibitemStop [0]{}%
\providecommand \bibitemNoStop [0]{.\EOS\space}%
\providecommand \EOS [0]{\spacefactor3000\relax}%
\providecommand \BibitemShut  [1]{\csname bibitem#1\endcsname}%
\let\auto@bib@innerbib\@empty
%</preamble>
\bibitem [{\citenamefont {Mermin}(2007)}]{Mm1}%
  \BibitemOpen
  \bibfield  {author} {\bibinfo {author} {\bibfnamefont {N.~D.}\ \bibnamefont {Mermin}},\ }\href@noop {} {\emph {\bibinfo {title} {Quantum Computer Science: An Introduction}}},\ \bibinfo {edition} {1st}\ ed.\ (\bibinfo  {publisher} {Cambridge University Press},\ \bibinfo {address} {Cambridge},\ \bibinfo {year} {2007})\BibitemShut {NoStop}%
\bibitem [{\citenamefont {Preskill}(2018)}]{Preq}%
  \BibitemOpen
  \bibfield  {author} {\bibinfo {author} {\bibfnamefont {J.}~\bibnamefont {Preskill}},\ }\href@noop {} {\bibfield  {journal} {\bibinfo  {journal} {Quantum}\ }\textbf {\bibinfo {volume} {2}},\ \bibinfo {pages} {1} (\bibinfo {year} {2018})}\BibitemShut {NoStop}%
\bibitem [{\citenamefont {Cerezo}\ \emph {et~al.}(2021{\natexlab{a}})\citenamefont {Cerezo}, \citenamefont {Arrasmith}, \citenamefont {Babbush}, \citenamefont {Benjamin}, \citenamefont {Endo}, \citenamefont {Fujii}, \citenamefont {McClean}, \citenamefont {Mitarai}, \citenamefont {Yuan}, \citenamefont {Cincio},\ and\ \citenamefont {Coles}}]{vqa}%
  \BibitemOpen
  \bibfield  {author} {\bibinfo {author} {\bibfnamefont {M.}~\bibnamefont {Cerezo}}, \bibinfo {author} {\bibfnamefont {A.}~\bibnamefont {Arrasmith}}, \bibinfo {author} {\bibfnamefont {R.}~\bibnamefont {Babbush}}, \bibinfo {author} {\bibfnamefont {S.~C.}\ \bibnamefont {Benjamin}}, \bibinfo {author} {\bibfnamefont {S.}~\bibnamefont {Endo}}, \bibinfo {author} {\bibfnamefont {K.}~\bibnamefont {Fujii}}, \bibinfo {author} {\bibfnamefont {J.~R.}\ \bibnamefont {McClean}}, \bibinfo {author} {\bibfnamefont {K.}~\bibnamefont {Mitarai}}, \bibinfo {author} {\bibfnamefont {X.}~\bibnamefont {Yuan}}, \bibinfo {author} {\bibfnamefont {L.}~\bibnamefont {Cincio}}, \ and\ \bibinfo {author} {\bibfnamefont {P.~J.}\ \bibnamefont {Coles}},\ }\href@noop {} {\bibfield  {journal} {\bibinfo  {journal} {Nature Communications}\ }\textbf {\bibinfo {volume} {3}},\ \bibinfo {pages} {625–644} (\bibinfo {year} {2021}{\natexlab{a}})}\BibitemShut {NoStop}%
\bibitem [{\citenamefont {Albino}\ \emph {et~al.}(2023)\citenamefont {Albino}, \citenamefont {Bloot},\ and\ \citenamefont {Gomes}}]{Anton}%
  \BibitemOpen
  \bibfield  {author} {\bibinfo {author} {\bibfnamefont {A.~S.}\ \bibnamefont {Albino}}, \bibinfo {author} {\bibfnamefont {R.}~\bibnamefont {Bloot}}, \ and\ \bibinfo {author} {\bibfnamefont {R.~F.~I.}\ \bibnamefont {Gomes}},\ }\href@noop {} {\bibfield  {journal} {\bibinfo  {journal} {Quantum Information Processing}\ }\textbf {\bibinfo {volume} {22}},\ \bibinfo {pages} {233} (\bibinfo {year} {2023})}\BibitemShut {NoStop}%
\bibitem [{\citenamefont {Cortes}\ and\ \citenamefont {Vapnik}(1995)}]{csvm}%
  \BibitemOpen
  \bibfield  {author} {\bibinfo {author} {\bibfnamefont {C.}~\bibnamefont {Cortes}}\ and\ \bibinfo {author} {\bibfnamefont {V.}~\bibnamefont {Vapnik}},\ }\href@noop {} {\bibfield  {journal} {\bibinfo  {journal} {Machine Learning}\ }\textbf {\bibinfo {volume} {20}},\ \bibinfo {pages} {273} (\bibinfo {year} {1995})}\BibitemShut {NoStop}%
\bibitem [{\citenamefont {Jerbi}\ \emph {et~al.}(2023)\citenamefont {Jerbi}, \citenamefont {Fiderer}, \citenamefont {Kübler}, \citenamefont {Briegel},\ and\ \citenamefont {Dunjko}}]{Jerbi}%
  \BibitemOpen
  \bibfield  {author} {\bibinfo {author} {\bibfnamefont {S.}~\bibnamefont {Jerbi}}, \bibinfo {author} {\bibfnamefont {H.}~\bibnamefont {Fiderer}, \bibfnamefont {Lukas J.and Poulsen~Nautrup}}, \bibinfo {author} {\bibfnamefont {J.~M.}\ \bibnamefont {Kübler}}, \bibinfo {author} {\bibfnamefont {H.~J.}\ \bibnamefont {Briegel}}, \ and\ \bibinfo {author} {\bibfnamefont {V.}~\bibnamefont {Dunjko}},\ }\href@noop {} {\bibfield  {journal} {\bibinfo  {journal} {Nat Commun}\ }\textbf {\bibinfo {volume} {14}},\ \bibinfo {pages} {517} (\bibinfo {year} {2023})}\BibitemShut {NoStop}%
\bibitem [{\citenamefont {Jumper}\ \emph {et~al.}(2021)\citenamefont {Jumper}, \citenamefont {Evans}, \citenamefont {Pritzel}, \citenamefont {Green}, \citenamefont {Figurnov}, \citenamefont {Ronneberger}, \citenamefont {Tunyasuvunakool}, \citenamefont {Bates}, \citenamefont {Žídek}, \citenamefont {Potapenko}, \citenamefont {Bridgland}, \citenamefont {Meyer}, \citenamefont {Kohl}, \citenamefont {Ballard}, \citenamefont {Cowie}, \citenamefont {Romera-Paredes}, \citenamefont {Nikolov}, \citenamefont {Jain}, \citenamefont {Adler}, \citenamefont {Back}, \citenamefont {Petersen}, \citenamefont {Reiman}, \citenamefont {Clancy}, \citenamefont {Zielinski}, \citenamefont {Steinegger}, \citenamefont {Pacholska}, \citenamefont {Berghammer}, \citenamefont {Bodenstein}, \citenamefont {Silver}, \citenamefont {Vinyals}, \citenamefont {Senior}, \citenamefont {Kavukcuoglu}, \citenamefont {Kohli},\ and\ \citenamefont {Hassabis}}]{Jump2021}%
  \BibitemOpen
  \bibfield  {author} {\bibinfo {author} {\bibfnamefont {J.}~\bibnamefont {Jumper}}, \bibinfo {author} {\bibfnamefont {R.}~\bibnamefont {Evans}}, \bibinfo {author} {\bibfnamefont {A.}~\bibnamefont {Pritzel}}, \bibinfo {author} {\bibfnamefont {T.}~\bibnamefont {Green}}, \bibinfo {author} {\bibfnamefont {M.}~\bibnamefont {Figurnov}}, \bibinfo {author} {\bibfnamefont {O.}~\bibnamefont {Ronneberger}}, \bibinfo {author} {\bibfnamefont {K.}~\bibnamefont {Tunyasuvunakool}}, \bibinfo {author} {\bibfnamefont {R.}~\bibnamefont {Bates}}, \bibinfo {author} {\bibfnamefont {A.}~\bibnamefont {Žídek}}, \bibinfo {author} {\bibfnamefont {A.}~\bibnamefont {Potapenko}}, \bibinfo {author} {\bibfnamefont {A.}~\bibnamefont {Bridgland}}, \bibinfo {author} {\bibfnamefont {C.}~\bibnamefont {Meyer}}, \bibinfo {author} {\bibfnamefont {S.~A.~A.}\ \bibnamefont {Kohl}}, \bibinfo {author} {\bibfnamefont {A.~J.}\ \bibnamefont {Ballard}}, \bibinfo {author} {\bibfnamefont {A.}~\bibnamefont {Cowie}}, \bibinfo {author} {\bibfnamefont
  {B.}~\bibnamefont {Romera-Paredes}}, \bibinfo {author} {\bibfnamefont {S.}~\bibnamefont {Nikolov}}, \bibinfo {author} {\bibfnamefont {R.}~\bibnamefont {Jain}}, \bibinfo {author} {\bibfnamefont {J.}~\bibnamefont {Adler}}, \bibinfo {author} {\bibfnamefont {T.}~\bibnamefont {Back}}, \bibinfo {author} {\bibfnamefont {S.}~\bibnamefont {Petersen}}, \bibinfo {author} {\bibfnamefont {D.}~\bibnamefont {Reiman}}, \bibinfo {author} {\bibfnamefont {E.}~\bibnamefont {Clancy}}, \bibinfo {author} {\bibfnamefont {M.}~\bibnamefont {Zielinski}}, \bibinfo {author} {\bibfnamefont {M.}~\bibnamefont {Steinegger}}, \bibinfo {author} {\bibfnamefont {M.}~\bibnamefont {Pacholska}}, \bibinfo {author} {\bibfnamefont {T.}~\bibnamefont {Berghammer}}, \bibinfo {author} {\bibfnamefont {S.}~\bibnamefont {Bodenstein}}, \bibinfo {author} {\bibfnamefont {D.}~\bibnamefont {Silver}}, \bibinfo {author} {\bibfnamefont {O.}~\bibnamefont {Vinyals}}, \bibinfo {author} {\bibfnamefont {A.~W.}\ \bibnamefont {Senior}}, \bibinfo {author} {\bibfnamefont
  {K.}~\bibnamefont {Kavukcuoglu}}, \bibinfo {author} {\bibfnamefont {P.}~\bibnamefont {Kohli}}, \ and\ \bibinfo {author} {\bibfnamefont {D.}~\bibnamefont {Hassabis}},\ }\href@noop {} {\bibfield  {journal} {\bibinfo  {journal} {Nature}\ }\textbf {\bibinfo {volume} {596}},\ \bibinfo {pages} {583} (\bibinfo {year} {2021})}\BibitemShut {NoStop}%
\bibitem [{\citenamefont {Rebentrost}\ \emph {et~al.}(2014)\citenamefont {Rebentrost}, \citenamefont {Mohseni},\ and\ \citenamefont {Loyd}}]{Loid}%
  \BibitemOpen
  \bibfield  {author} {\bibinfo {author} {\bibfnamefont {P.}~\bibnamefont {Rebentrost}}, \bibinfo {author} {\bibfnamefont {M.}~\bibnamefont {Mohseni}}, \ and\ \bibinfo {author} {\bibfnamefont {S.}~\bibnamefont {Loyd}},\ }\href@noop {} {\bibfield  {journal} {\bibinfo  {journal} {Physical Review Letters}\ }\textbf {\bibinfo {volume} {113}},\ \bibinfo {pages} {130503} (\bibinfo {year} {2014})}\BibitemShut {NoStop}%
\bibitem [{\citenamefont {Havlíček}\ \emph {et~al.}(2019)\citenamefont {Havlíček}, \citenamefont {Córcoles}, \citenamefont {Temme}, \citenamefont {Harrow}, \citenamefont {Kandala}, \citenamefont {Chow},\ and\ \citenamefont {Gambetta}}]{qsvm}%
  \BibitemOpen
  \bibfield  {author} {\bibinfo {author} {\bibfnamefont {V.}~\bibnamefont {Havlíček}}, \bibinfo {author} {\bibfnamefont {A.~D.}\ \bibnamefont {Córcoles}}, \bibinfo {author} {\bibfnamefont {K.}~\bibnamefont {Temme}}, \bibinfo {author} {\bibfnamefont {A.}~\bibnamefont {Harrow}}, \bibinfo {author} {\bibfnamefont {A.}~\bibnamefont {Kandala}}, \bibinfo {author} {\bibfnamefont {J.}~\bibnamefont {Chow}}, \ and\ \bibinfo {author} {\bibfnamefont {J.}~\bibnamefont {Gambetta}},\ }\href@noop {} {\bibfield  {journal} {\bibinfo  {journal} {Nature Communications}\ }\textbf {\bibinfo {volume} {567}},\ \bibinfo {pages} {209} (\bibinfo {year} {2019})}\BibitemShut {NoStop}%
\bibitem [{\citenamefont {Liu}\ \emph {et~al.}(2021{\natexlab{a}})\citenamefont {Liu}, \citenamefont {Arunachalam},\ and\ \citenamefont {Temme}}]{Yunchao}%
  \BibitemOpen
  \bibfield  {author} {\bibinfo {author} {\bibfnamefont {Y.}~\bibnamefont {Liu}}, \bibinfo {author} {\bibfnamefont {S.}~\bibnamefont {Arunachalam}}, \ and\ \bibinfo {author} {\bibfnamefont {K.}~\bibnamefont {Temme}},\ }\href@noop {} {\bibfield  {journal} {\bibinfo  {journal} {Nat Physics}\ }\textbf {\bibinfo {volume} {17}},\ \bibinfo {pages} {1013–1017} (\bibinfo {year} {2021}{\natexlab{a}})}\BibitemShut {NoStop}%
\bibitem [{\citenamefont {McClean1}\ \emph {et~al.}(2018)\citenamefont {McClean1}, \citenamefont {Boixo}, \citenamefont {Smelyanskiy}, \citenamefont {Babbush1},\ and\ \citenamefont {Neven}}]{Jarrod}%
  \BibitemOpen
  \bibfield  {author} {\bibinfo {author} {\bibfnamefont {J.~R.}\ \bibnamefont {McClean1}}, \bibinfo {author} {\bibfnamefont {S.}~\bibnamefont {Boixo}}, \bibinfo {author} {\bibfnamefont {V.~N.}\ \bibnamefont {Smelyanskiy}}, \bibinfo {author} {\bibfnamefont {R.}~\bibnamefont {Babbush1}}, \ and\ \bibinfo {author} {\bibfnamefont {H.}~\bibnamefont {Neven}},\ }\href@noop {} {\bibfield  {journal} {\bibinfo  {journal} {Nat Commun}\ }\textbf {\bibinfo {volume} {9}},\ \bibinfo {pages} {4812} (\bibinfo {year} {2018})}\BibitemShut {NoStop}%
\bibitem [{\citenamefont {Cerezo}\ \emph {et~al.}(2021{\natexlab{b}})\citenamefont {Cerezo}, \citenamefont {Sone}, \citenamefont {Volkoff1}, \citenamefont {Cincio},\ and\ \citenamefont {Coles}}]{cerezoP}%
  \BibitemOpen
  \bibfield  {author} {\bibinfo {author} {\bibfnamefont {M.}~\bibnamefont {Cerezo}}, \bibinfo {author} {\bibfnamefont {A.}~\bibnamefont {Sone}}, \bibinfo {author} {\bibfnamefont {T.}~\bibnamefont {Volkoff1}}, \bibinfo {author} {\bibfnamefont {L.}~\bibnamefont {Cincio}}, \ and\ \bibinfo {author} {\bibfnamefont {P.~J.}\ \bibnamefont {Coles}},\ }\href@noop {} {\bibfield  {journal} {\bibinfo  {journal} {Nat Commun}\ }\textbf {\bibinfo {volume} {12}},\ \bibinfo {pages} {1791} (\bibinfo {year} {2021}{\natexlab{b}})}\BibitemShut {NoStop}%
\bibitem [{\citenamefont {Schuld}\ and\ \citenamefont {Killoran}(2022)}]{Schud2022}%
  \BibitemOpen
  \bibfield  {author} {\bibinfo {author} {\bibfnamefont {M.}~\bibnamefont {Schuld}}\ and\ \bibinfo {author} {\bibfnamefont {N.}~\bibnamefont {Killoran}},\ }\href@noop {} {\bibfield  {journal} {\bibinfo  {journal} {PRX Quantum}\ }\textbf {\bibinfo {volume} {3}},\ \bibinfo {pages} {030101} (\bibinfo {year} {2022})}\BibitemShut {NoStop}%
\bibitem [{\citenamefont {Schuld}\ and\ \citenamefont {Petruccione}(2021)}]{schudbook}%
  \BibitemOpen
  \bibfield  {author} {\bibinfo {author} {\bibfnamefont {M.}~\bibnamefont {Schuld}}\ and\ \bibinfo {author} {\bibfnamefont {F.}~\bibnamefont {Petruccione}},\ }\href@noop {} {\emph {\bibinfo {title} {Quantum Models as Kernel Methods}}}\ (\bibinfo  {publisher} {Springer},\ \bibinfo {year} {2021})\BibitemShut {NoStop}%
\bibitem [{\citenamefont {Schuld}\ and\ \citenamefont {Killoran}(2021)}]{Schud2021}%
  \BibitemOpen
  \bibfield  {author} {\bibinfo {author} {\bibfnamefont {M.}~\bibnamefont {Schuld}}\ and\ \bibinfo {author} {\bibfnamefont {N.}~\bibnamefont {Killoran}},\ }\href@noop {} {\bibfield  {journal} {\bibinfo  {journal} {Phys. Rev. A}\ }\textbf {\bibinfo {volume} {103}},\ \bibinfo {pages} {032430} (\bibinfo {year} {2021})}\BibitemShut {NoStop}%
\bibitem [{\citenamefont {Macaluso}\ \emph {et~al.}(2023)\citenamefont {Macaluso}, \citenamefont {Klusch}, \citenamefont {Lodi},\ and\ \citenamefont {Sartori}}]{Malacuso}%
  \BibitemOpen
  \bibfield  {author} {\bibinfo {author} {\bibfnamefont {A.}~\bibnamefont {Macaluso}}, \bibinfo {author} {\bibfnamefont {M.}~\bibnamefont {Klusch}}, \bibinfo {author} {\bibfnamefont {S.}~\bibnamefont {Lodi}}, \ and\ \bibinfo {author} {\bibfnamefont {C.}~\bibnamefont {Sartori}},\ }\href@noop {} {\bibfield  {journal} {\bibinfo  {journal} {Quantum Information Processing}\ }\textbf {\bibinfo {volume} {22}},\ \bibinfo {pages} {159} (\bibinfo {year} {2023})}\BibitemShut {NoStop}%
\bibitem [{\citenamefont {Gyurik}\ \emph {et~al.}(2023)\citenamefont {Gyurik}, \citenamefont {Dyon~Vreumingen},\ and\ \citenamefont {Dunjko}}]{Casper}%
  \BibitemOpen
  \bibfield  {author} {\bibinfo {author} {\bibfnamefont {C.}~\bibnamefont {Gyurik}}, \bibinfo {author} {\bibfnamefont {v.}~\bibnamefont {Dyon~Vreumingen}}, \ and\ \bibinfo {author} {\bibfnamefont {V.}~\bibnamefont {Dunjko}},\ }\href@noop {} {\bibfield  {journal} {\bibinfo  {journal} {Quantum}\ }\textbf {\bibinfo {volume} {7}},\ \bibinfo {pages} {893} (\bibinfo {year} {2023})}\BibitemShut {NoStop}%
\bibitem [{\citenamefont {F.Araujo}\ \emph {et~al.}(2021)\citenamefont {F.Araujo}, \citenamefont {Park}, \citenamefont {Petruccione},\ and\ \citenamefont {da~Silva}}]{Araujo}%
  \BibitemOpen
  \bibfield  {author} {\bibinfo {author} {\bibfnamefont {I.}~\bibnamefont {F.Araujo}}, \bibinfo {author} {\bibfnamefont {D.~K.}\ \bibnamefont {Park}}, \bibinfo {author} {\bibfnamefont {F.}~\bibnamefont {Petruccione}}, \ and\ \bibinfo {author} {\bibfnamefont {A.~J.}\ \bibnamefont {da~Silva}},\ }\href@noop {} {\bibfield  {journal} {\bibinfo  {journal} {Nat Scientifc Reports}\ }\textbf {\bibinfo {volume} {11}},\ \bibinfo {pages} {6329} (\bibinfo {year} {2021})}\BibitemShut {NoStop}%
\bibitem [{\citenamefont {Mitarai}\ \emph {et~al.}(2018)\citenamefont {Mitarai}, \citenamefont {Negoro}, \citenamefont {Kitagawa},\ and\ \citenamefont {Fujii}}]{Circlern0}%
  \BibitemOpen
  \bibfield  {author} {\bibinfo {author} {\bibfnamefont {K.}~\bibnamefont {Mitarai}}, \bibinfo {author} {\bibfnamefont {M.}~\bibnamefont {Negoro}}, \bibinfo {author} {\bibfnamefont {M.}~\bibnamefont {Kitagawa}}, \ and\ \bibinfo {author} {\bibfnamefont {K.}~\bibnamefont {Fujii}},\ }\href@noop {} {\bibfield  {journal} {\bibinfo  {journal} {Phys. Rev. A}\ }\textbf {\bibinfo {volume} {98}},\ \bibinfo {pages} {032309} (\bibinfo {year} {2018})}\BibitemShut {NoStop}%
\bibitem [{\citenamefont {Franken}\ \emph {et~al.}(2020)\citenamefont {Franken}, \citenamefont {Georgiev}, \citenamefont {Mücke}, \citenamefont {Wolter}, \citenamefont {Heese}, \citenamefont {Bauckhage},\ and\ \citenamefont {Piatkowski}}]{qce}%
  \BibitemOpen
  \bibfield  {author} {\bibinfo {author} {\bibfnamefont {L.}~\bibnamefont {Franken}}, \bibinfo {author} {\bibfnamefont {B.}~\bibnamefont {Georgiev}}, \bibinfo {author} {\bibfnamefont {S.}~\bibnamefont {Mücke}}, \bibinfo {author} {\bibfnamefont {M.}~\bibnamefont {Wolter}}, \bibinfo {author} {\bibfnamefont {R.}~\bibnamefont {Heese}}, \bibinfo {author} {\bibfnamefont {C.}~\bibnamefont {Bauckhage}}, \ and\ \bibinfo {author} {\bibfnamefont {N.}~\bibnamefont {Piatkowski}},\ }\href {\doibase 10.48550/ARXIV.2012.13453} {\enquote {\bibinfo {title} {Quantum circuit evolution on nisq devices},}\ } (\bibinfo {year} {2020})\BibitemShut {NoStop}%
\bibitem [{\citenamefont {Blum}\ and\ \citenamefont {Reymond}(2009)}]{qm91}%
  \BibitemOpen
  \bibfield  {author} {\bibinfo {author} {\bibfnamefont {L.~C.}\ \bibnamefont {Blum}}\ and\ \bibinfo {author} {\bibfnamefont {J.-L.}\ \bibnamefont {Reymond}},\ }\href@noop {} {\bibfield  {journal} {\bibinfo  {journal} {J. Am. Chem. Soc.}\ }\textbf {\bibinfo {volume} {131}},\ \bibinfo {pages} {8732} (\bibinfo {year} {2009})}\BibitemShut {NoStop}%
\bibitem [{\citenamefont {Rupp}\ \emph {et~al.}(2012)\citenamefont {Rupp}, \citenamefont {Tkatchenko}, \citenamefont {M\"uller},\ and\ \citenamefont {von Lilienfeld}}]{qm92}%
  \BibitemOpen
  \bibfield  {author} {\bibinfo {author} {\bibfnamefont {M.}~\bibnamefont {Rupp}}, \bibinfo {author} {\bibfnamefont {A.}~\bibnamefont {Tkatchenko}}, \bibinfo {author} {\bibfnamefont {K.-R.}\ \bibnamefont {M\"uller}}, \ and\ \bibinfo {author} {\bibfnamefont {O.~A.}\ \bibnamefont {von Lilienfeld}},\ }\href@noop {} {\bibfield  {journal} {\bibinfo  {journal} {Physical Review Letters}\ }\textbf {\bibinfo {volume} {108}},\ \bibinfo {pages} {058301} (\bibinfo {year} {2012})}\BibitemShut {NoStop}%
\bibitem [{\citenamefont {Valdés}\ and\ \citenamefont {Tchagang}(2023)}]{qm9_1}%
  \BibitemOpen
  \bibfield  {author} {\bibinfo {author} {\bibfnamefont {J.~J.}\ \bibnamefont {Valdés}}\ and\ \bibinfo {author} {\bibfnamefont {A.~B.}\ \bibnamefont {Tchagang}},\ }\href@noop {} {\enquote {\bibinfo {title} {Understanding the structure of qm7b and qm9 quantum mechanical datasets using unsupervised learning},}\ } (\bibinfo {year} {2023}),\ \Eprint {http://arxiv.org/abs/2309.15130} {arXiv:2309.15130 [physics.chem-ph]} \BibitemShut {NoStop}%
\bibitem [{\citenamefont {Liu}\ \emph {et~al.}(2021{\natexlab{b}})\citenamefont {Liu}, \citenamefont {Arunachalam},\ and\ \citenamefont {Temme}}]{speedupsupervised}%
  \BibitemOpen
  \bibfield  {author} {\bibinfo {author} {\bibfnamefont {Y.}~\bibnamefont {Liu}}, \bibinfo {author} {\bibfnamefont {S.}~\bibnamefont {Arunachalam}}, \ and\ \bibinfo {author} {\bibfnamefont {K.}~\bibnamefont {Temme}},\ }\href {\doibase 10.1038/s41567-021-01287-z} {\bibfield  {journal} {\bibinfo  {journal} {Nature Physics}\ }\textbf {\bibinfo {volume} {17}},\ \bibinfo {pages} {1745} (\bibinfo {year} {2021}{\natexlab{b}})}\BibitemShut {NoStop}%
\bibitem [{\citenamefont {Hjorungnes}\ and\ \citenamefont {Gesbert}(2007)}]{dermat}%
  \BibitemOpen
  \bibfield  {author} {\bibinfo {author} {\bibfnamefont {A.}~\bibnamefont {Hjorungnes}}\ and\ \bibinfo {author} {\bibfnamefont {D.}~\bibnamefont {Gesbert}},\ }\href {\doibase 10.1109/TSP.2007.893762} {\bibfield  {journal} {\bibinfo  {journal} {IEEE Transactions on Signal Processing}\ }\textbf {\bibinfo {volume} {55}},\ \bibinfo {pages} {2740} (\bibinfo {year} {2007})}\BibitemShut {NoStop}%
\bibitem [{\citenamefont {Fisher}(1936)}]{Fisher}%
  \BibitemOpen
  \bibfield  {author} {\bibinfo {author} {\bibfnamefont {R.~A.}\ \bibnamefont {Fisher}},\ }\href@noop {} {\bibfield  {journal} {\bibinfo  {journal} {Annals of Eugenics}\ }\textbf {\bibinfo {volume} {7}},\ \bibinfo {pages} {179} (\bibinfo {year} {1936})}\BibitemShut {NoStop}%
\end{thebibliography}%
\end{document}